# PulsatioMech: An Open-Source MATLAB Toolbox for Seismocardiography Signal Processing

**Nathan Zavanelli**
George W. Woodruff School of Mechanical Engineering,
College of Engineering,
Georgia Institute of Technology Atlanta, GA 30332, USA
nzavanelli@gatech.edu

## I. Introduction

This paper summarizes and presents PulsatioMech: an open-source MATLAB toolbox for seismocardiography (SCG) signal processing. The toolbox may be found here: https://github.com/nzavanelli/SCG_master_toolbox
PulsatioMech is currently under development as a common tool to promote new studies and discoveries in the use of cardiac mechanical signal for wearable health monitoring. This toolbox is designed to assist users in analyzing SCG signals without the need to devote significant effort into signal processing and coding tasks. Simultaneously, it provides a uniform basis to assess the reproducibility of works based on this toolbox, including those cited here [1-6]. The referenced works contain a great deal more detail regarding the specific algorithms implemented here, whereas this paper will present a short overview of the PulsatioMech Toolbox.

## II. Toolbox contents

This toolbox is subdivided into 8 constituent folders. These are:

1. Acc

A set of functions dedicated to analyzing the gross acceleration in an SCG signal. This acceleration is low pass filtered to only capture movements in the single Hz range, and the outputs are primarily used for respiratory, position, and actigraphy analysis.

2. Beat-averaging

This folder is dedicated to ensembling and averaging SCG beats to provide information about a set of beats over time. Tools for rejecting noisy beats and basic classification are also provided.

3. Filtering

This folder contains functions used to filter SCG signals.

4. MATLAB-fex

This folder contains a diverse set of helper functions used to process SCG signals.

5. Plotting

This folder contains several useful plotting functions.

6. SCG

This folder holds the majority of the useful SCG specific functions that constitute the PulsatioMech toolbox. Specific details about each of the functions may be found in the methods sections of the works cited in this paper.

7. Signal-quality

This folder contains several methods for assessing the signal quality of non-sinusoidal signals, like SCG. Although these algorithms were designed with SCG in mind, they are broadly applicable to numerous biosignals, like electrocardiography (ECG) and photoplethysmography (PPG).

8. Spectral

This folder contains several functions pertaining to the spectral analysis of SCG signals.

## III. Feature generation overview

One of the central innovations in this toolbox is the introduction of multifractal analysis and other advanced signal processing techniques for SCG signals. These include five broad categories, and the following description is largely adapted from the following work [6]. The first category of features are autoregressive coefficients derived via Burg's method, which were chosen to act as a surrogate for autocorrelation periodicity of the signal over time. The next set of features corresponds to detail coefficients of various wavelet models of the signal based on the Daubechies wavelet, which well captures scale-dependent morphological signal changes. The Shannon entropy based on a maximal overlap discrete wavelet packet transform with the length 18 Fejér-Korovkin wavelet of order four is also introduced as a way to capture the overall kinetic energy in a stressed and relaxed beat, which has been shown to change with physiological arousal (Morra et al. 2021). Finally, significant utility is derived from the estimation of how multifractal each beat is with respect to both wavelet scales and moments. Every fractal system has a fractal dimension h (referred to as a Holder exponent) defined by:

$$\exists h \geq 0 \, s.t.$$
$$|f(x) - f(y)| \sim \|x - y\|^h$$
$$\forall x, y \in dom \, f$$

Here, h is determined for each leading wavelet scale. If there is a wide range of h, then the system has a wide spread of fractal dimensions and is thus termed multifractal. A stressed beat is noticeably less multifractal than a relaxed beat (as shown by the lesser spread in Holder exponents). It can also be shown that the magnitude of each scaling coefficient associated with these exponents is linear as a function of moment for a purely mono-fractal system, and the degree of non-linearity infers a more multifractal process. This analysis is performed for each beat, showing that relaxed beats are again more multifractal.

More details regrading each method are provided here and reproduced exactly from [6].

*Multifractal 1-D wavelet leader estimation*
The wavelet leaders were determined based on the following procedure:
- Calculate the wavelet coefficients, $d_x(j,k)$, from the discrete wavelet transform and determine the absolute value of each coefficient for all scales. Note that each successive finer scale has twice the number of coefficients. Determine the dyadic interval at scale $2^j$ as a union of two intervals at a finer scale as shown below.

$$\left[2^j k, 2^j(k+1)\right) = \left[2^{j-1}(2k), 2^{j-1}(k+2)\right)$$
$$[2^{j-1}(2k), 2^{j-1}(2k+2))$$
$$= [2^{j-1}(2k), 2^{j-1}(2k+1))$$
$$\cup [2^{j-1}(2k+1), 2^{j-1}(k+2))$$

- Begin with the scale one level coarser than the finest obtained scale.
- Compare the above value to all its finer dyadic intervals and note the maximum value.
- Repeat for the next value and continue along the scale.
- From the maximum values obtained for a given scale, examine the first three values and obtain the maximum of those neighbors. That maximum value is a leader for that scale.
- Continue comparing the maximum values to obtain the other leaders for that scale.
- Progress to the next coarser scale and repeat.

*Global Hölder exponent estimates*
The Hölder exponents, which quantify the local regularity, are determined from the wavelet leader maxima. The singularity spectrum indicates the size of the set of Hölder exponents in the data. The scaling coefficients ζ(q) may be estimated as:

$$\frac{1}{n_a}\sum_{k=1}^{n_a} |T_x(a,k)|^q = a^{\zeta(q)}$$

where a is the scale, q is the moment, Tx are the wavelet leaders by scale, $n_a$ is the number of wavelet leaders at each scale, and ζ(q) is the scaling exponent.
Expanding ζ(q) to a polynomial produces:
$$\zeta(q)=c_1 q+c_2 q^2/2+c_3 q^3/6+\ldots$$

*Shannon entropy (SE) values*
The following includes steps to calculate the SE values for the maximal overlap discrete wavelet packet transform (MODPWT) at level 4:

1) Select the length of the 18 Fejér-Korovkin wavelet of order 4.
2) Decompose the original signal according to the specified W and L.
3) Calculate the energy in each coefficient for each node in the finest level L (terminal node)
4) Calculate the energy probability distribution for each coefficient in each terminal node
5) Calculate the entropy of the terminal node
These entropy values form the Shannon entropy coefficient vector.

*Autoregressive reflection coefficients based on Burg's method*
Given an input signal $x_0, x_0, \ldots, x_{N-1}$, the m reflection coefficients may be determined by the following approach:
1) Initialize $f'_0 = b'_0 = [x_0, x_0, \ldots, x_{N-1}]^T$ and iteration i=0.
2) Remove the first element of $f'_i$ and last element of $b'_i$
$$f_i = f'_i(1:N-i-1)$$
$$b_i = b'_i(0:N-i-2)$$
3) Calculate the reflection coefficient k as:
$$k_i = -\frac{2b_i^H f_i}{f_i^H f_i + b_i^H b_i}$$
4) Update the prediction errors
$$f_{i+1} = f_i + k_i b_i$$
$$b_{i+1} = b_i + k_i^* f_i$$
5) Increment i = I + 1
6) Return to step 2.

IV. APPLICATIONS

SCG is an infrequently explored, yet powerful tool for assessing the heart's mechanical function. When the heart beats, it produces a mechanical vibration that manifests in longitudinal, planar and transverse waves that propagate through the chest tissue. These waves combine and manifest on the chest wall as the SCG signal. Based on existing seismology tools, one can analyze these vibrations to ascertain information about the heart's mechanical function, including fiducial timings, kinetic energy, and beat profile. This information can be potentially useful in identifying sympathetic innervations that affect cardiac timings, abnormal beat morphologies, total cardiac output, blood pressure, and the presence of cardiovascular disease. Significant work remains to be done in creating a robust and uniform understanding of the connection between observed beats and fundamental cardiac mechanics and capturing SCG signals during periods of motion.

V. TOOLBOX DEMONSTRATION

This toolbox includes a simple demonstration of the PulsatioMech algorithms and a sample SCG signal. In this demonstration, the SCG signal is loaded and filtered. It is then ensemble averaged to produce a representative average beat for a given time period. A template beat is then produced against which subsequent beats can be compared. The difference between these next beats and the template is then computed. A waterfall plot is then produced to show how the SCG morphology changes over time. Finally, the five categories of advanced features are then produced, and the multifractal analysis plots are provided.

## VI. OPEN SCIENCE

All source code for this toolbox is available as an open source project under an MIT License.

## VII. GETTING INVOLVED

Additional researchers are highly encouraged to contribute to the toolbox. Any bug reports or source code contributions should be submitted via merge requests to the repository, and detailed guidelines regarding contributions can be found on the GitLab repository pages. Users should code in MATLAB 2018a or higher.

## VIII. ACKNOWLEDGEMENTS

The author would like to thank all the colleagues and former developers who have contributed to the PulsatioMech toolbox and those who will do so in the future. The author would also like to thank the Georgia Tech Institute for Electronics and Nanotechnology (IEN) and the National Science Foundation Graduate Research Program (NSF GRFP) for their support.

## IX. REFEENCES